\documentclass[comsoc,journal]{IEEEtran} 
\usepackage{comment}
\usepackage{cite,graphicx,amsmath,amssymb}
\usepackage{graphicx,caption}
\usepackage{amsfonts,amssymb}
\usepackage{subfigure}
\usepackage{xcolor}
\usepackage{multirow}
\usepackage{url}
\usepackage{enumerate}
\usepackage[
top    = 1.70cm,
bottom = 0.82 in,
left   = 0.63 in,
right  = 0.63 in]{geometry}
\usepackage{xspace}
\newcommand\ahmad[1]{{\textcolor{black}{#1}}}
\newcommand\jiadongnew[1]{{\textcolor{black}{#1}}}
\begin{document}

\title{6G Mobile-Edge Empowered Metaverse: Requirements, Technologies, Challenges and Research Directions}
\author{Jiadong~Yu $^*$, Ahmad~Alhilal $^*$, Pan~Hui, Danny~H.K.~Tsang
%\thanks{ J. Yu is with the Hong Kong University of Science and Technology (Guangzhou).}
%\thanks{A. Alhilal is with the Hong Kong University of Science and Technology.} \\
\thanks{\textbf{ *~J. Yu and A. Alhilal contributed equally to this work~}}
%\thanks{P. Hui and  D.H.K. Tsang are with the Hong Kong University of Science and Technology (Guangzhou), and with the Hong Kong University of Science and Technology.}
}

\maketitle

\begin{abstract}
The Metaverse has emerged as the successor of the conventional mobile internet to change people’s lifestyles. It has strict visual and physical requirements to ensure an immersive experience (i.e., high visual quality, low motion-to-photon latency, and real-time tactile and control experience). 
However, the current technologies fall short to satisfy these requirements.
Mobile edge computing (MEC) has been indispensable to enable low latency and powerful computing. Moreover, the sixth generation (6G) networks promise to provide end users with seamless communications.
In this paper, we explore and demonstrate the synergistic relationship between 6G and mobile-edge technologies in empowering the Metaverse with ubiquitous communications and computation. 
This includes the usage of heterogeneous radios, intelligent reflecting surfaces (IRS), non-orthogonal multiple access (NOMA), and digital twins (DTs) - assisted MEC.
We also discuss emerging communication paradigms (i.e., semantic communication, holographic-type communication, and haptic communication) to further satisfy the demand for human-type communications and fulfill user preferences and immersive experiences in the Metaverse.
\end{abstract}

\section{Introduction}
%Metaverse 
The Metaverse is the blended space at the intersection between physical and virtual spaces, extended reality (XR). XR includes many facets, specifically, Virtual Reality (VR), Augmented Reality (AR), Mixed Reality (MR), and tactile internet (TI). It is regarded as the successor to the conventional Internet, changing the way people work, entertain, and socialize. 
%reviewer 1 comment 5
%With the widespread of the internet of things (IoT) and embedded sensors, data can be collected to create digital twins (DTs) in the virtual world. Conversely, DTs can assist in physical decision-making using simulation and machine learning. 
Immersive Metaverse applications require high-capacity wireless communication to deliver voluminous data with low latency. Although basic XR can be supported by 5G, this technology falls short to support the widespread use of XR devices. In Release 18 (5G-Advanced), the Third Generation Partnership Project (3GPP) explores empowering the XR experience via New Radio (NR)-based communications, which is covered in the work item -- Study on XR enhancements for NR. NR includes two frequency ranges that span 410 MHz to 7125 MHz and 24250 MHz to 71000 MHz.  In its recent technical report~\cite{3gpp_xr}, 3GPP studies enhancements for multi-modality service, interaction networks among communicating entities, QoS and policy enhancements for media service transmission, and power management. Further 3GPP works (i.e., Release 18+) are expected to provide capacity and power enhancements in support of advanced media services (e.g., AR, VR, and XR).

% \jiadong{In its recent technical report, re support of XR services, the technical report\cite{3gpp_xr} discussed within 3GPP technical specification group Services and System Aspects studying the enhancements for multi-modality service, the network for interaction, QoS and policy enhancements for media service transmission, and power management. It is anticipated that further 3GPP activities and work items related to providing capacity and power enhancements for XR services will be scheduled in later releases (i.e., 18+) on top of the findings of the ongoing study in Release 18.}
%current 3GPP-REL-18, potential towards REL-19

%Emerging 6G technologies can eliminate such physical barriers to enable the desired experience in the Metaverse.

In this paper, we provide an overview of the Metaverse and its requirements and investigate the enabling 6G technologies and communication paradigms, specifically, from the perspective of how these technologies can empower the Metaverse to fulfill human-centric preferences and immersive experience. This contribution is expected to align with 3GPP efforts related to XR in 3GPP Release 19, specifically the ongoing work item -- Study on Localized Mobile Metaverse Services. Potentially, work items related to new paradigms for the Metaverse can be considered.
 % \jiadong{that can potentially be further discussed in the coming 3GPP Release 19. Specifically, from the perspective that how these technologies can empower Metaverse to fulfill human-centric preferences and immersive experience}.
 We then discuss some challenges that may arise when deploying 6G technologies, as well as the research directions to cope with them. Naturally, the following three research questions (RQs) arise:

\textbf{RQ 1: What are the requirements for a high quality of service/experience (QoS/E) in the Metaverse?}
\ahmad{With the widely used devices (i.e., head-mounted devices (HMDs), AR glasses, and gloves) to access Metaverse, the QoS/E can be defined based on visual and physical aspects.}
Visually, VR systems demand a bitrate up to 1 Gbps to deliver a video streaming of real-like views. Besides, motion-to-photon (MTP) latencies should be less than 20 ms to provide smooth movements in the virtual space
~\cite{8329628}. 
Physically, the XR system should eventually allow people to control and steer both real and virtual objects. 
Control and tactile experiences would require real-time transmission and \jiadongnew{high-reliability} feedback of haptic information, especially for time-sensitive applications.
\begin{figure*}[!t]
  \centering
  \includegraphics[width=.75\textwidth]{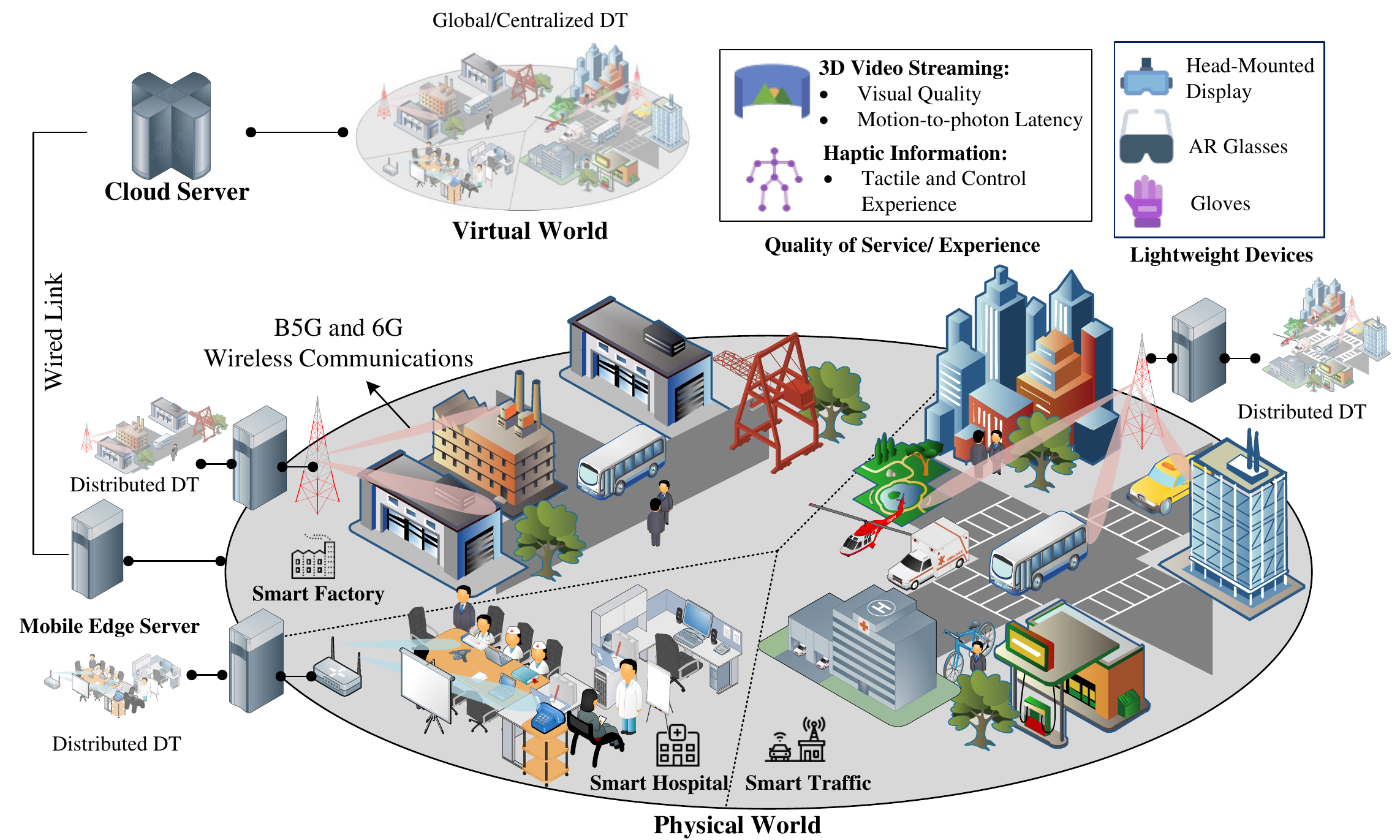}\\
  \caption{6G Mobile-edge empowered Metaverse.}
  \label{Metaverse}
\end{figure*}

\textbf{\jiadongnew{RQ 2: What is the role of 6G technologies and MEC for high QoS/E in the Metaverse?}}
% \jiadongnew{There is a synergistic relationship between 6G and MEC to empower the Metaverse, which resides in their ability to provide ubiquitous connectivity and computation. This combination enables high QoS/E by ensuring seamless connectivity and offloading intensive computing tasks to the edge servers. 6G networks leveraging techniques such as heterogeneous radios and intelligent reflecting surfaces (IRS) for ubiquitous connectivity and non-orthogonal multiple access (NOMA) for spectral efficiency to provide high-reliability and low-latency connections across various environments. This ensures that users can access the Metaverse from anywhere, whether they are indoors, outdoors, or on the move. Additionally, MEC complements 6G by extending computational capabilities to the edge of the network. The processing power required for complex computations can be shifted from the user's device to the edge server. Combining DT technology with MEC enables tracking resource utilization, latency, throughput, and other performance metrics. This helps in optimizing the resource allocation of the MEC system to improve the immersive user experience in the Metaverse.}
\ahmad{While 6G can provide high-capacity and low-latency communications %to MEC servers
between users and MEC servers
, MEC can provide computing and caching capabilities close to users. Therefore, the synergy of 6G and MEC features provides ubiquitous connectivity and computation. 
These features ensure seamless connectivity and heavy task offloading to the MEC servers, enabling data-intensive and real-time applications in the Metaverse and providing high QoS/E. 
%Besides, 6G can provide low energy consumption which enables heavy metaverse applications to run on lightweight devices such as head-mounted displays (HDMs). 
6G networks achieve high-reliable and low-latency connections across various environments by leveraging techniques such as heterogeneous radios and intelligent reflecting surfaces (IRS) for ubiquitous connectivity and non-orthogonal multiple access (NOMA) for spectral efficiency. These technologies allow users to access the Metaverse anywhere (e.g., indoors, outdoors, or on the move) at any time.} 
\jiadongnew{Additionally, MEC complements 6G by moving complex computations from the device to the network edge.}
\ahmad{Incorporating DT technology into MEC enables monitoring of the MEC system's resources, user's end-to-end latency, throughput, and other performance metrics. This helps in optimizing the resource allocation of the MEC system to improve the Metaverse user experience.}

%Metaverse systems enable the high-quality rendering of virtual scenes on lightweight clients (e.g., head-mounted displays and mobile devices) by offloading such intensive computing tasks to mobile edge computing (MEC) servers. Frames of rendered scenes are then streamed to the clients through mobile networks. However, Metaverse applications such as VR 360° video streaming are bandwidth-hungry and latency-sensitive. With the increasing of Metaverse users, the fourth (4G) and fifth generations (5G), which are based on sub-6 frequency bands, fall short to satisfy the bandwidth demand for high-quality video streaming. The sixth generation (6G) can combine  The mobile-edge systems can take advantage of these strategies to provide ubiquitous connectivity and high data rates. However, communications using mmWave and THz require a line of sight (LoS) between the client and the access point.
%Review 1 comment 8
%\jiadong{When obstacles block the LoS, intelligent reflecting surfaces (IRS) can tune the communication beams towards the user by constructing the virtual LoS links}, thus enhancing communication reliability. The physical objects and users (i.e., physical twins -- PTs) have digital replicas (i.e., digital twins -- DTs) in the Metaverse. DTs are synchronized with PTs and contain historic and recent data of their physical counterparts, allowing MEC to train artificial intelligence (AI) models for optimal resource allocation.

\textbf{RQ 3: What are the emerging human-type communications for non-traditional user experience in the Metaverse?}
%Reviewer1 comment 1
%With the increasing demand for intelligent services in the Metaverse, 6G communications is anticipated to shift from a traditional architecture that focuses solely on a high transmission rate to a new architecture based on the intelligent connection of everything.
With the growing demand for intelligent services in the Metaverse, a 6G communication architecture is essential to provide intelligent connection of everything. This architecture will differ from traditional architecture, which only emphasizes a high transmission rate.
As a revolutionary architecture, semantic communication integrates user preference, application requirements, and the meaning of information into data processing and transmission. This architecture helps to eliminate the transmission of redundant information and reach beyond Shannon communications.
The high-speed communication in the 6G networks would enable holographic-type communication. Reproducing holograms of physical individuals through digital projections would take XR to the next level. Additionally, haptic systems supplement the audiovisual experience with sensation in the Metaverse.
% The high-speed communication in the 6G networks would ensure holographic-type communication becomes a new communication scenario. The reproducing holograms of physical individuals through digital projections would take XR to a new level. Additionally, haptic systems extend the audiovisual experience with sensation in the Metaverse.}

\section{Metaverse Requirements}
\label{sec:requirements}
%\subsection{Metaverse}
In this section, we identify the QoS/E requirements in the Metaverse and discuss the necessity to deploy 6G and MEC.
\subsection{Quality of Service/Experience (QoS/E)}
As illustrated in Fig.~\ref{Metaverse}, the Metaverse is the post-reality universe, XR, that allows people to work, play, and interact in an immersive three-dimensional (3D) virtual environment. In the Metaverse, physical and virtual worlds are blended into mixed physical-virtual environments. Such environments generate an imaginary environment similar to the real world, including realistic sounds, images, and other sensations~\cite{lee2021all}. 
%The Metaverse has strict requirements for a fully-immersive experience, large-scale concurrent users, and seamless connectivity, which poses unprecedented challenges to the communication system, such as ubiquitous connectivity, ultra-low latency, and ultra-high capacity and reliability~\cite{lee2021all}. 
\textbf{XR} is a broad category of reality that employs immersive technologies to create digital environments in which data are represented and projected. XR includes many facets, specifically VR, AR, and MR. Human users interact partially or fully with each other and with the mixed physical-virtual environment. 
%\textbf{VR} platforms (e.g., Mozilla Hub, Spatial, and Meta Workrooms) create separate digital and artificial environments in which users feel immersed and behave as they would in real life. \textbf{AR} overlays digital objects into the physical environment to provide more engaging information to the users. It spatially augments the physical layer with virtual objects using mobile devices such as smartphones, tablets, glasses, contact lenses, or other transparent surfaces. \textbf{MR} is an advanced version of AR that liberates users from screen-bound experiences by offering instinctual interactions with data, users, and virtual objects in living spaces. Supported by strong environmental comprehension or situational awareness capabilities, MR objects can coexist with other real objects in physical environments.
% \jiadongnew{In the Metaverse, there are mainly three types of lightweight immersive equipment that serve as the primary means of interacting with virtual environments. These include HMDs, AR glasses, and gloves Haptic Devices (shown in Fig.~\ref{Metaverse}). While HMDs and AR glasses primarily focus on the visual aspects of immersion, gloves, and haptic devices contribute to the tactile and physical interactions in the Metaverse. These devices work together from both visual and physical perspectives to create a more immersive (i.e., visual quality and tactile and control experience) and interactive (i.e., MTP latency) experience for users, blurring the boundaries between the physical and virtual worlds.}
\ahmad{In the Metaverse, there are mainly three types of lightweight immersive equipment that serve as means of interaction with virtual environments. These include HMDs, AR glasses, and gloves with haptic sensors (shown in Fig.~\ref{Metaverse}). While HMDs and AR glasses primarily handle the visual aspect of immersion, gloves and haptic devices ensure tactile and physical interaction. The combination of visual and physical experiences using these devices can bring more immersion (i.e., visual quality and tactile and control experience). Besides, the integration of edge computing can improve the interaction (i.e., lower MTP latency), blurring the boundaries between the physical and virtual worlds.} 
% Latency requirement

%The Metaverse is intended to provide satisfactory user experiences that include frames rendered at high-quality, view-user interaction (MTP latency), and real-time tactile experience for immersion from both visual and physical perspectives. 
\textbf{Visual Quality:} Human eyes resolution can reach up to 64 M pixels (150° horizontal and 120° vertical field-of-view (FOV), 60 pixels/degree) at a certain moment. A reality-like view requires 120 frames per second. Assuming a colored pixel is represented in 4 bytes and considering a maximum video compression ratio of 1:600 (using H.265 HEVC codec), the VR system would demand a bitrate of up to 1 Gbps to deliver such video quality~\cite{8329628}. 
\textbf{MTP Latency} is the time elapsed between a movement (e.g., head motion, rotation, and pitch) and the display of the corresponding FOV to the human eyes. High MTP latency values send contradictory signals to the vestibulo-ocular reflex (VOR), which can cause motion sickness. This latency should be less than 20 ms to provide smooth movements in the virtual space~\cite{8329628}.
\textbf{Tactile and Control Experience:}
The Metaverse should allow people to control real and virtual objects \jiadongnew{with high or ultra-high reliability}, and provide them with real-time control and physical tactile experiences. In smart hospitals, for instance, telesurgery should allow surgeons to feel the sensation of the touch and force feedback via haptic clothing/equipment. Therefore, implementation of the control actions and adjustments is essential. The communication channel must ensure real-time transmission of haptic information (i.e., touch, actuation, motion, vibration, and surface texture)~\cite{9681620}.
\subsection{\jiadongnew{Synergized 6G and MEC Empowered Metaverse}}
% \jiadongnew{The synergy between 6G and the MEC lies in their combined ability to provide the seamless communication infrastructure and computing capabilities to support the Metaverse. 
% Mobile edge computing, also known as the edge computing infrastructure, brings computational capabilities and storage closer to the end-users, reducing latency and enabling real-time processing and decision-making. As illustrated in Fig.~\ref{Metaverse}, Metaverse users primarily use lightweight devices such as HMD, AR glasses, gloves, and other wearable devices. These devices rely on mobile networks, including B5G and 6G wireless communications, to access the extensive array of services offered by the Metaverse. This integration ensures a seamless and omnipresent experience for users across both indoor and outdoor environments.
\ahmad{The synergy of 6G and MEC ensures seamless connectivity and ubiquitous computing which empowers the Metaverse. 6G, the next generation of wireless technology, aims to provide ultra-high-reliability, ultra-low-latency, and ubiquitous connectivity, enabling transformative applications and services. MEC brings computational capabilities and storage closer to the end-users, reducing latency and enabling real-time data processing and decision-making.
Fig.~\ref{Metaverse} illustrates three use cases of the synergy between 6G and MEC in the Metaverse, smart factory, smart hospital, and smart traffic. }\jiadongnew{The Metaverse presents demanding requirements and constantly generates resource-intensive tasks for DT generation (e.g., image processing, data processing) and service access (e.g., FOV prediction and rendering).
Metaverse users wear light devices (e.g., HMD, AR glasses, and gloves) that connect to MEC services via mobile networks over B5G and 6G wireless communications. This enables a smooth and ubiquitous experience in any environment. The devices offload heavy tasks to MEC servers due to their inherent constraints of the devices (e.g., limited resources and battery life). }

\section{\jiadongnew{6G Mobile-Edge} Empowered Metaverse}
%\section{6G Technologies for Mobile-Edge Empowered Metaverse}
In this section, we discuss the main technology components in the 6G mobile-edge-empowered Metaverse framework. 
%We first investigate wireless communication technologies and bidirectional DT and MEC in the Metaverse. Moreover, we provide an evaluation of a \jiadongnew{6G-mobile edge case study}. We introduce new communication paradigms that can enrich the user experience. 
Table \ref{table} summarizes the main 6G mobile-edge technologies that can assist in improving different aspects of QoS/E in the Metaverse. 

\begin{table}
\centering
\caption{Summary of enabling 6G mobile-edge technologies.}
\label{table}
\begin{tabular}{|l|l|l|l|l|} 
\hline
\multicolumn{2}{|l|}{\multirow{2}{*}{\parbox{3.5cm}{\centering \textbf{Main 6G mobile-edge technologies}}}}                                       & \multicolumn{3}{l|}{\parbox{3.5cm}{\centering \textbf{QoS/E for the Metaverse}}}                        \\ 
\cline{3-5}
\multicolumn{2}{|l|}{}                                                                              & \parbox{1cm}{\centering Visual quality} & \parbox{1cm}{\centering MTP latency} & \parbox{1cm}{\vspace{.25\baselineskip}\centering Tactile and control\vspace{.25\baselineskip}}   \\ 
\hline
\hline
\multirow{3}{*}{\parbox{1.5cm}{\centering \textbf{Wireless communication technologies}\vspace{.25\baselineskip}}} &  \parbox{2cm}{\centering \vspace{.25\baselineskip}Heterogeneous radios\vspace{.25\baselineskip}}              & \checkmark           & \checkmark                     & \checkmark                            \\ 
\cline{2-5}
                                                                  & \parbox{2cm}{\centering \vspace{.25\baselineskip} IRS\vspace{.25\baselineskip}}             & \checkmark           & \checkmark                     & \checkmark                            \\ 
\cline{2-5}
                                                                  & \parbox{2cm}{\centering \vspace{.25\baselineskip}NOMA\vspace{.25\baselineskip}}                        & \checkmark           & \checkmark                     & \checkmark                            \\
\hline                                                            \multirow{2}{*}{\parbox{1.5cm}{\centering \textbf{Bidirectional DT and MEC}}}                            & \parbox{2cm}{\centering \vspace{.25\baselineskip}MEC-enabled DT\vspace{.25\baselineskip}}                     & \checkmark           & \checkmark                     & \checkmark                            \\ 
\cline{2-5}
                                                                  & \parbox{2cm}{\centering \vspace{.25\baselineskip}DT-assisted MEC\vspace{.25\baselineskip}}                     & \checkmark           & \checkmark                     & \checkmark                            \\

\hline

\multirow{3}{*}{\parbox{1.5cm}{\centering \textbf{New communication paradigms}}} & \parbox{2cm}{\centering \vspace{.25\baselineskip}Semantic communication\vspace{.25\baselineskip}}                      &                & \checkmark                     & \checkmark                            \\ 
\cline{2-5}
                                                                  & \parbox{2cm}{\centering \vspace{.25\baselineskip}Holographic-type communication\vspace{.25\baselineskip}} & \checkmark           &                     &                                 \\ 
\cline{2-5}
                                                                  & \parbox{2cm}{\centering\vspace{.25\baselineskip} Haptic communication\vspace{.25\baselineskip}}    &                &                          & \checkmark                            \\ 
\hline
\end{tabular}
\end{table}

\subsection{Communication Technologies \jiadongnew{for Wireless VR}}
With the widespread adoption of the Metaverse, 5G communications aimed to achieve the 20Gb/s peak data transmission target will not be sufficient. 6G intelligent networks are anticipated to combine ground-breaking technologies to attain a capacity of at least 1 Tb/s and peak data rates of 10 Tb/s.
%\cite{8766143}.
\subsubsection{Heterogeneous Radios}
\label{sec:het-radios}
\begin{figure*}[t]
  \centering
  \includegraphics[width=.95\textwidth]{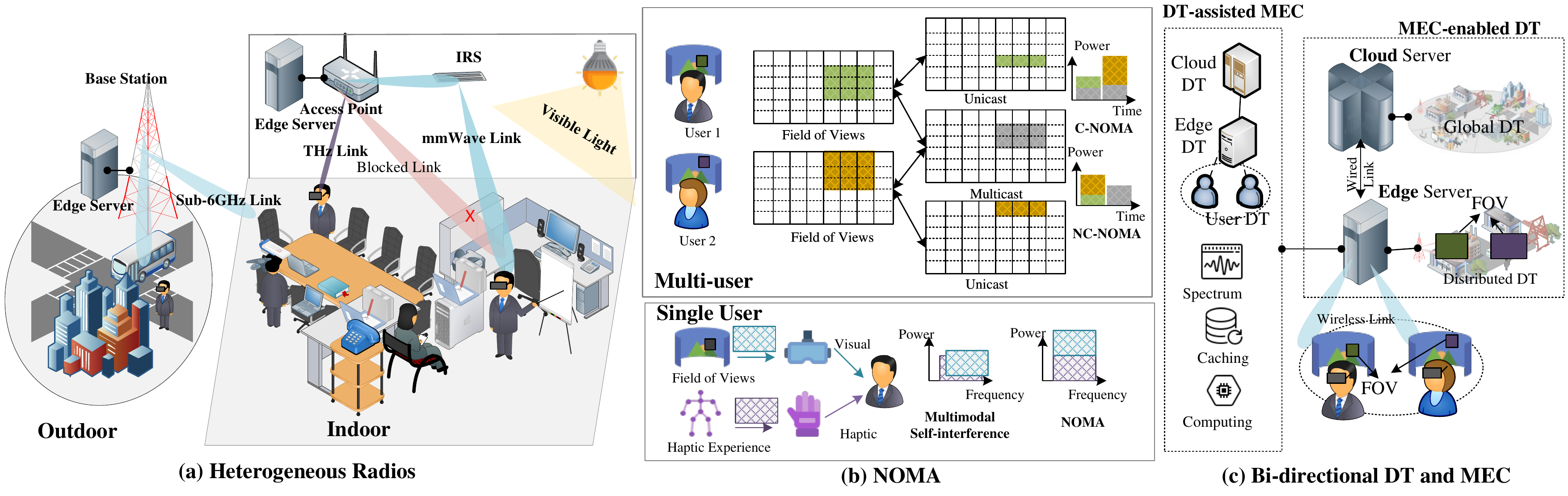}\\
  \caption{Wireless communication technologies and bidirectional DT and MEC for the Metaverse.}
  \label{wireless}
\end{figure*}
Current commercial 5G wireless networks utilize the sub-6 GHz spectrum (3.4 GHz - 3.8 GHz) to provide wide signal coverage. However, the explosive growth of online intelligent services leads to a spectral deficiency. 
With the high capacity of mmWave spectrum, utilizing mmWave frequency bands ($<$ 50  GHz) is expected to increasingly dominate in 5G communication systems. However, mmWave communications encounter high path loss due to signal blockage, including the case of being blocked by users' hands. 
Separately caching heterogeneous videos such as monocular and stereoscopic videos in two-tier heterogeneous networks (dual-connectivity sub-6GHz and mmWave base stations) can enhance the reliability in Metaverse while ensuring high capacity\cite{9536410}.

Although mmWave beams can focus on a tiny area with the aid of beamforming, managing interference between nearby users in a crowded environment remains a challenge. THz frequencies (95 GHz-3 THz) have wide bandwidth with vast potential for voluminous data streaming. They are foreseen as natural competitors to deliver an unprecedented high data rate in the Metaverse. THz beamforming can spatially construct and align narrower pencil beams to Metaverse users and hence considerably reduce interference, compared to mmWave. However, narrow pencil beamforming in dense indoor VR scenarios would require passively adjusting the beam direction or user association after the user moves. This can detach the users from their virtual world. Visible light communication (VLC), a frequency band between 430 THz to 750 THz, based on light-emitting diodes (LEDs) can provide an alternative and accurate positioning service. Wang et al~\cite{9745789} design a framework, in which LEDs provide precise indoor locating services for users via VLC, while small base stations (SBSs) broadcast high-quality Metaverse content to users through THz bands. The framework's THz/VLC-enabled wireless VR network can jointly consider the VLC access points (VAPs) selection and user association in order to provide reliable positioning and a high data rate for VR content transmission to the end users.

Fig.~\ref{wireless}(a) illustrates the coexistence of multiple radio bands. In 6G wireless empowered Metaverse, both indoor and outdoor users can change between the frequency bands (e.g., sub-6  GHz, mmWave,  THz, and VLC), enabling heterogeneous radio networks, or can simultaneously utilize multiple bands using heterogeneous network interfaces. This makes it possible to link users to the network as a whole in the cell-less network protocols and take advantage of the features of each band. Additionally, it will support seamless mobility without consuming extra communication overhead caused by handovers between access points and/or base stations.

\subsubsection{Intelligent Reflecting Surface}
\label{sec:irs}

Fig.~\ref{wireless}(a) demonstrates a use case integration of multiple bands (mmWave, THz, and sub-6 GHz) for indoor and outdoor communications. mmWave and  THz signals mostly work in LoS which is subject to obstruction. The communication succeeds only when the transmitter's beam and the receiver are aligned. Any slight misalignment can lead to halting the data stream from the Metaverse services. 
%To cope with the misalignment, different from conventional full-duplex relays that amplify and forward the signal of interests without directionality, 
%mmWave mirror captures the mmWave signal and reflects in the direction of the users~\cite{abari2017enabling}. In contrast to full-duplex relays which require complex analog and digital hardware with full transmit and receive chains, the mmWave mirror features only an analogue front-end (i.e., antennae and an amplifier).
IRS has recently been extensively studied by the communication community. IRS can reflect the impinging signal towards the target direction. Differently, IRS consists of numerous passive reflecting elements which can smartly reconfigure the wireless propagation environment by tuning the phase shifts. 
Chaccour et al~\cite{9149411} utilize  THz, IRS, and MEC to empower the VR systems in confined indoor areas. This ensures high data rates and reliable low latency for seamless user connectivity.

\subsubsection{Non-orthogonal Multiple Access}
\label{sec:noma}
Conventional orthogonal multiple access (OMA) techniques such as orthogonal frequency division multiple access (OFDMA), can serve a single user in an orthogonal spectrum resource block. In contrast to OMA, NOMA can serve multiple users in each orthogonal resource block, sharing the same time/frequency/code domain. NOMA applies different power levels to meet the rising demand for high spectral efficiency and capacity. It uses successive interference cancellation (SIC) and superposition coding (SC) in its modulation and demodulation. The deployment of NOMA would provide significant advantages for bandwidth-hungry Metaverse applications. In live VR viewing scenarios, for instance, utilizing behavioral coherence can improve the transmission reliability of the jointly requested data and overall capacity. Multiple users might request virtual content (e.g., VR 3D streaming) at the same time with overlapping users' FOVs, and some tiles of FOVs are shared. Using NOMA, the content delivery system in the Metaverse can use multicast to transmit the shared FOV tiles, and unicast to transmit the non-shared tiles.  
The multi-user case of Fig.~\ref{wireless}(b) illustrates two users with overlapping FOVs and two transmission schemes, cooperative-NOMA (C-NOMA) and non-cooperative-NOMA (NC-NOMA). These schemes are utilized in \cite{9120553}, exploiting additional temporal diversity gains. 
In C-NOMA, the transmission reliability is increased by employing two-time slots to transmit the jointly requested data (shared FOV tiles). These slots are overlaid by the requested data of individual users (non-shared FOV tiles).
In NC-NOMA, however, the non-shared FOV tiles are superposed and transmitted in the first time slot to the requested users, whereas the shared FOV tiles are transmitted in the second time slot using OMA. 

In addition, as shown in the single-user case of Fig. 2(c), NOMA can assist in concurrently transmitting visual content (FOV tiles) and haptic senses to individual users~\cite{8647208}. The visual perception requires intermediate reliability, maximum possible rate, and packet error rate on the order of $10^{-1}$ to $10^{-3}$, whereas the haptic perception requires a constant rate, ultra-high reliability, and packet error rate on the order of $10^{-4}$ to $10^{-5}$. Therefore, visuo-haptic VR traffic necessitates the usage of two distinct network services: enhanced mobile broadband (eMBB) for visual perception and ultra-reliable and low latency communication (URLLC) for haptic perception. However, the spectrum sharing between multimodal transmission of the eMBB and URLLC services poses self-interference issues (i.e., multimodal self-interference). NOMA scheme with reliable SIC can assist to overcome this issue and help the Metaverse system achieve a higher haptic data rate as well as a lower MTP latency.
\subsection{Bidirectional DT and MEC \jiadongnew{to empower the Metaverse}}
\ahmad{A DT is a primary component in the Metaverse, defined as the real-time evolving digital replica of a physical object, process, or system that spans its lifecycle.} Real-time communications are essential to synchronize DTs with their PTs. Any state updates in DTs reflect on their PTs, and vice versa. AI supported by DTs can assist to improve the Metaverse performance, level of efficiency, and accuracy.
With the explosive deployment of the IoT, the use of DTs is prevailing in applications such as smart traffic, smart factory, and smart hospital (see Fig.~\ref{Metaverse}). In the smart factory, for instance, the AI-supported DT can provide real-time input, and forecast the performance of the production line and machine status. This enables the manufacturer to anticipate issues before they happen and boosts the effectiveness of the production process. %Just like a sizable manufacturing facility that can be handled by tuning many settings on their DTs, 
Likewise, the MEC ecosystem can be managed and troubleshot to enhance its performance using its DT~\cite{9696282}. Therefore, there is bidirectional reliance between DT and MEC in the Metaverse.

\subsubsection{MEC-enabled DT}
\ahmad{DTs can be operated on the edge servers located in close proximity to the physical system. This allows for faster and low-latency data processing, as well as real-time feedback and decision-making based on the collected data.} As mentioned above, AI algorithms can rely on the DT to predict how the physical process will perform using up-to-date and historical data. However, training AI models require large datasets and can be very computation-intensive~\cite{9696282}. As illustrated in Fig.~\ref{Metaverse} and Fig.~\ref{wireless}(c), in the MEC-empowered Metaverse, maintaining DTs, performing rendering, and executing AI tasks can be offloaded to edge servers. Additionally, the distributed DTs stored at edge servers can be shared with a central server (cloud) for constructing a global DT. This global DT represents a digital replica of the entire physical world. 

\subsubsection{DT-assisted MEC}
\ahmad{DT-assisted MEC is the DT representation of the edge computing system.} As illustrated in Fig.~\ref{wireless}(c), this DT can reflect the dynamics of the MEC ecosystem. It can monitor the most up-to-date state of edge computing, resource utilization, and network performance in real-time. It can also fabricate emulation of users, load, and resource allocation to produce emulated data. These data help to train the DT model to find the optimal strategy for resource allocation. Accordingly, DT-assisted MEC can make decisions and send feedback to the edge computing system to assign the resources to the users that ensure optimal QoE and fairness QoE. In particular, DTs provide historical and real-time data of the MEC ecosystem, based on which, AI can train models to generate strategies. These strategies can, for instance, automatically switch the operating frequency band to achieve the target bandwidth (see Section~\ref{sec:het-radios}), adjust the IRS phase shifts to align the users' directions (see Section~\ref{sec:irs}), allocate levels of power for VR data transmission in NOMA (see Section~\ref{sec:noma}. Moreover, DT-assisted MEC can also continuously monitor the condition of the whole system, detect the unfairness in resource distribution, or forecast any system failure in edge computing, thus enabling proactive intervention.

\begin{figure}[t]
  \centering
  \includegraphics[width=.45\textwidth]{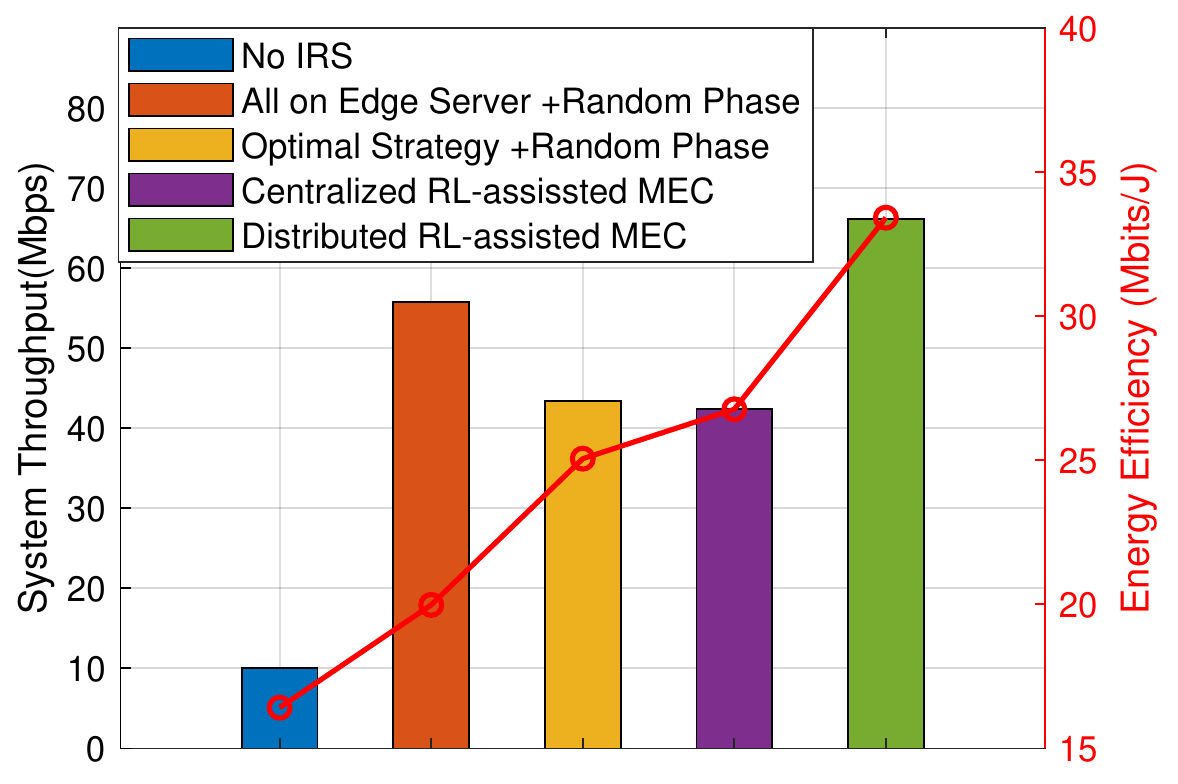}
  \vspace{-.2cm}
   % \caption{System performance of IRS-assisted NOMA-aided \jiadongnew{MEC system with RL-supported algorithm.}}
    \caption{ \ahmad{Performance comparison of RL variants for resource allocation over mobile-edge system.}} %Distributed RL over a 6G Mobile-edge system (with IRS and NOMA) achieves the highest throughput and energy efficiency.}}
   
  \label{results}
\end{figure}
\subsection{Case Study Evaluation of 6G Mobile-edge}

\ahmad{Since the users' devices accessing the Metaverse have limited battery capacity and are lightweight, we focus our case study on evaluating the energy efficiency and system throughput of a 6G mobile-edge system. Optimizing energy efficiency is crucial to ensure prolonged device operation and an uninterrupted user experience. The system is required to handle the substantial data demands of the Metaverse’s diverse applications and services that entail high-quality video streaming, real-time rendering, and interactive communication. As such, we also evaluate the network throughput of the 6G mobile-edge system to meet such demands. }To verify the utility of the aforementioned communication technologies and DT-assisted MEC, we design a model emulating a MEC system using Matlab. The model integrates IRS and NOMA and uses reinforcement learning (RL) for energy-efficient allocation and high throughput communication.

We assume tasks can be executed locally on users' devices and on edge servers. The size of IRS elements is $64$, and there are $4$ users in the NOMA sub-carrier cluster. Details about the system and configurations can be found in~\cite{9968198}. Centralized RL-supported MEC refers to the agent deployment at the edge server, whereas distributed DT-assisted MEC refers to the heterogeneous agents deployment on both edge server and user devices. The MEC system utilizes centralized RL Deep Deterministic Policy Gradient (DDPG) and distributed RL multi-agent DDPG (MADDPG) to find the optimal strategy for allocating resources. Without IRS, the system exhibits the worst energy efficiency and lowest throughput, as shown in Fig.~\ref{results}. Centralized RL enhances energy efficiency by approximately 63\%, compared to Without IRS. While distributed RL enhances the energy efficiency further by 40\%, compared to the centralized RL. Although executing all the computation tasks on MEC servers can achieve higher system throughput, this solution degrades the energy efficiency by more than 40\%, depleting the system batteries rapidly, compared to distributed RL. As a result, with distributed RL-supported MEC, the optimal strategy for offloading decisions, power allocation for NOMA, and IRS phase shifts can be obtained, thus maximizing the utility of the MEC system. %Therefore, this prototype can be deployed in the Metaverse to fulfill the stringent requirements discussed in Section~\ref{sec:requirements}.  
\subsection{\jiadongnew{New Communication Paradigms}}

The above-mentioned technologies would improve the QoS/E in the Metaverse. Unconventional new communication paradigms that are based on user preference and application scenarios (i.e., semantic, holographic-style, and haptic communication), are needed to efficiently use limited communication and computation resources.

\subsubsection{Semantic Communication}
%Using IoT, user and sensor data can be collected from user devices to create DTs in the Metaverse. Conversely, activities and interactions in the virtual world (e.g., collected via IoT actuation) have an impact on the physical domain. Future communication systems must be capable of transmitting a large volume of data to create a detailed and high-quality virtual world. This world replicates precisely the physical world in real-time. 
The conventional communication system aims to reach the highest data transmission rates and the smallest symbol (bit) error rate. However, even with advanced encoding, decoding, and modulation techniques, the system capacity is approaching the Shannon physical capacity limit.
To overcome this limitation, semantic communication, an emerging communication paradigm,
captures semantic features and uses AI to extract the meanings of the transmitted information and reconstruct data at the target.
%Since the receiver concerns only about the meaning of transmitted messages rather than the accurate bit stream, the semantic features of raw source data can be extracted and encoded at the transmitter before transmission. 
%Although there could be syntactic inconsistencies on the receiver side, there are no semantic mistakes. This implies that a semantic communication system can still function well and consume less energy when bandwidth is limited or the signal-to-noise ratio (SNR) is quite low.  
For real-time interaction in the Metaverse, the encoder at the user side learns the semantic representations of the multimodal data (i.e., visual and haptic data), while the generator at the receiver side (i.e., the edge server) learns how to manipulate them for scene rendering and interactions\cite{park2022enabling}. Only semantic features will be transmitted (instead of raw data) which reduces the required communication resources. Thus, it increases communication efficiency, reducing the MTP latency and improving the control and physical tactile experiences.

% communication  thus increasing communication efficiency\cite{park2022enabling}.
% By transmitting semantic representations instead of the raw data, the burden on wireless communications can be relieved and thus reducing the MTP latency and improving real-time control and physical tactile experiences for the Metaverse}. 

%\begin{figure}[!t]
%  \centering
%  \includegraphics[width=3.4in]{figures/newcom.pdf}\\
%  \caption{\jiadong{New communication paradigms for the Metaverse.}}
%  \label{newcom}
%\end{figure}
\subsubsection{Holographic-type Communication}
A 3D view in the Metaverse is rendered and displayed to the users using HMDs as a sequence of two-dimensional (2D) views of the same scene (left and right eyes). Different from that, the hologram adds parallax which changes depending on where the viewer is looking, allowing him/her to interact with the projection~\cite{8970173}. 
The parallax instantly tricks the brain and creates an illusion of having a physical object or real environment mixed with the 3D digital overlay. 
On the transmitter side, the holographic system captures the users from various perspectives using an array of cameras. The omnidirectional photos are then merged, turned into a hologram, and encoded for streaming through the network. The transmitted hologram is decoded and rendered on the receiver side for a holographic display. Holograms will make it possible to, for instance, make eye contact with individuals in a meeting. The visual experience immediately evokes emotions and stimulates the senses, causing activity in the cortex to increase, allowing the viewer to process and retain the information communicated more effectively.
Holograms generate a digital projection of the physical individual rather than a modified or entirely fictitious portrayal of the person. %In the Metaverse, they are the ideal candidate to replace avatars.
Therefore, holographic-type communication is expected to improve the user's perceptual quality with visual preference when delivering the Metaverse content.

\subsubsection{Haptic Communication}
Haptic feedback complements the visual and audio to provide an immersive experience in the Metaverse. Haptic feedbacks encompass both kinesthetic perception (information about position, torques, velocity, and forces felt by tendons, joints, and muscles of the human body), and tactile perception (sensation of surface roughness and friction by various skin mechanoreceptors)~\cite{9681620}. 
%Fig.~\ref{newcom}(c) illustrates three applications of haptic feedback, remote surgery, robotic arm control, and driving.
Haptic Communication seeks to fulfill the requirements of high-quality of control and tactile experiences, especially, ultra-reliable and responsiveness to enable real-time human-to-machine interaction.
Defined by IEEE P1918.1 standards working group%~\footnote{\url{https://grouper.ieee.org/groups/1918/1/}}
, TI is a breakthrough network for remotely accessing, perceiving, manipulating, or controlling real and virtual objects or processes in real-time. TI is intended to shift the paradigm from content-oriented to steer/control-based communication enabling the real-time transmission of haptic information (e.g., touch, actuation, motion, vibration, and surface texture) along with audiovisual data over the Internet. The service and medium relationship between haptic communication and TI are comparable to that between voice communication over Internet protocol and the Internet.

%\subsubsection{\jiadongnew{Internet of Senses}}
%\jiadongnew{The Internet of Senses (IoS) encompasses more than just haptic feedback. While haptic feedback, which provides a sense of touch and tactile sensations, is an important aspect of the IoS, there are other sensory experiences that it aims to incorporate as well. IoS refers to an emerging concept that aims to extend the traditional Internet to incorporate sensory experiences beyond the scope of conventional audiovisual communication. It envisions a networked ecosystem where digital technologies and the physical world converge to enable immersive and multi-sensory interactions. The IoS encompasses various technologies and innovations that enable the transmission, manipulation, and perception of sensory information over the Internet. These sensory experiences can include not only visual and auditory stimuli but also other senses such as touch, taste, and smell. By integrating these additional senses into digital experiences, the IoS aims to create more immersive, realistic, and engaging interactions between users and the digital world\cite{panagiotakopoulos2022digital}. Some examples of IoS applications include VR and AR systems that incorporate odour-emitting devices that can transmit scents for enhanced virtual experiences; taste simulators that can recreate flavours digitally; and haptic feedback, enabling users to feel and touch virtual objects.}

\section{Challenges and Research Directions}
This section discusses some technical challenges that would encounter in the development and deployment of Metaverse applications. They are related to the integration of heterogeneous technologies, five-sense user experience, and AI generative content (AIGC). 
\subsection{Integration of Heterogeneous Technologies}
Integration of 6G technologies for a mobile-edge system can assist to achieve the required bandwidth for the Metaverse. However, this poses additional challenges to managing diverse communication resources and services, such as network congestion and the associated latency (higher latency due to queuing).
The DT-assisted MEC can use trained AI models to forecast or identify busy times and overload links when providing Metaverse services. Once network bottlenecks are discovered, alternate frequency bands can be requested for multi-carrier switching or merging%~\cite{Alhilal2022distri}
, and IRS reflection beams can be adjusted toward crowded areas. NOMA clusters can be intelligently assigned for multiple users or heterogeneous data from a single user.
%\subsubsection{DT-assisted MEC}
Although many emerging studies provide theoretical performance optimization for latency and transmission throughput in DT-assisted MEC, these works are neither implemented nor verified in real-world scenarios. Future studies should focus on deploying DT-assisted MEC prototypes, implementing the entire pipeline of data collection, and both direction PT-to-DT synchronization. They should also investigate the additional power consumption of the synchronization and the proper placement of DTs across the MEC system.
% \subsection{User-oriented Experience with Five Senses}
\subsection{\ahmad{Five-Sense User Experience}}
%challenges: five senses and fairness
\ahmad{ The sensory experiences may also include sensations like taste and smell in addition to the QoS/E listed in Table~\ref{table}. In the real world, people's preferences and sensitivities for sensory stimuli vary. Future Metaverse should consider the human's five senses\cite{panagiotakopoulos2022digital}. Therefore, the key challenge is accurately modeling user perception (i.e., individual preferences, sensitivities, and cognitive responses to different sensory stimuli) in the Metaverse to define QoE. Moreover, under the 6G mobile-edge empowered Metaverse framework, users in the same environment can share both the communication resources (i.e., spectrum) and the edge computation resources (i.e., caching memory and computation frequency). To avoid issues such as network congestion, system collapse, and user abandonment caused by the unfair resource allocation in the Metaverse, it is essential to satisfy fairness in QoE and resource allocation.
Research should also focus on designing user models and developing machine learning algorithms and adaptive systems that dynamically adjust five senses stimuli to meet individual preferences. Further study should investigate leveraging edge intelligence and AI algorithms for personalized experiences. This would support customized Metaverse services (e.g., visual content recommendations).}

\subsection{AI Generated Content (AIGC)}
\ahmad{
%challenges
AIGC refers to the way of content creation using artificial intelligence. AIGC involves the creation of digital content, such as images, music, and natural languages using AI models\cite{xu2023unleashing}. It can generate realistic virtual environments, characters, narratives, and other elements that enhance the overall Metaverse experience. However, generating and rendering such content in real-time  pose challenges due to the limited processing power and resources available on remote servers. Additionally, ensuring data privacy, security, and interoperability across different platforms and devices remains a challenge to be addressed. 
%direction
To handle the demands of multiple users and AIGC in the Metaverse, massive computational power for data processing and efficient data transmission in the 6G-mobile edge framework is needed. Research can also focus on developing privacy-preserving AI techniques, secure data-sharing protocols, and encryption mechanisms to safeguard user data from algorithms design, 6G communications, and edge computation perspectives.}

%\subsection{Privacy and Security}
%Reviewer1comment#4
%6G networks are designed to accommodate the internet of everything (IoE). Thus, IoE must allow decentralized systems to make intelligent decisions on various levels and scales.Constructing AI-supported distributed or centralized DT on both edge and cloud would require sensitive and private data from the users (e.g., physiological reactions and body motions). Therefore, privacy-preserving and security remain major issues.
%Using the distributed learning methodology such as the federated learning approach, the users' visual or haptic data collected by the lightweight devices can be trained locally, and the trained model parameters, which (compared to the raw data) are more difficult to extract insights, will be shared and aggregated in the MEC system, thus eliminating any privacy violations. In terms of security, hacked communication links or edge servers could have negative effects on Metaverse services. For example, attacks related to AI algorithms (i.e., data transmission, training, inference, or deployment attacks) may cause algorithm failure, leading to mismatch or disruptions of the Metaverse services. DT-assisted MEC could capture the users' activity, identify any deviation from normal behavior, and detect any anomalies. 

\section{Conclusion}
This article overviews the \ahmad{synergy of 6G and edge computing that provide ubiquitous communication and computation to empower the Metaverse.} 
6G networks integrate multiple technologies to provide wireless Metaverse, such as heterogeneous radios, IRS, and NOMA. The DT-assisted MEC can train AI models to assist in optimally allocating \ahmad{the communication and computation resources. Our evaluation results proved that 6G-mobile edge systems can improve energy efficiency and provide higher throughput, thus supporting the Metaverse application with lightweight devices.} 
Semantic communication, holographic-type communication, and haptic communication have emerged to fulfill a variety of human-type communications, user preferences, and non-conventional experiences in the Metaverse. Finally, this paper highlights some challenges that would arise when deploying a 6G mobile-edge framework and discussed research directions on aspects of integration of heterogeneous technologies, five-sense user experience, and AIGC.

\bibliographystyle{IEEEtran}
\bibliography{IEEEabrv,mybib}
\renewenvironment{IEEEbiography}[1]
  {\IEEEbiographynophoto{#1}}
  {\endIEEEbiographynophoto}
  \vskip -2.5\baselineskip plus -1fil
\begin{IEEEbiography}{Jiadong Yu} (jiadongyu@ust.hk) is currently a Postdoc Fellow at the Internet of Things thrust at the Hong Kong University of Science and Technology (Guangzhou). 
\end{IEEEbiography}
\vskip -2.5\baselineskip plus -1fil
\begin{IEEEbiography}{Ahmad Alhilal} (aalhilal@connect.ust.hk) is currently a Postdoc Fellow at the Hong Kong University of Science and Technology, IPO, Division of EMIA.
\end{IEEEbiography}
\vskip -2.5\baselineskip plus -1fil
\begin{IEEEbiography}{Pan Hui}(panhui@ust.hk) is a Chair Professor of Computational Media and Arts thrust at the Hong Kong University of Science and Technology (Guangzhou), and a Chair Professor of Emerging Interdisciplinary Areas and Director of the HKUST-DT Systems and Media Lab, Hong Kong University of Science and Technology.
\end{IEEEbiography}
\vskip -2.5\baselineskip plus -1fil
\begin{IEEEbiography}{Danny Hin-Kwok Tsang}(eetsang@ust.hk) is the Thrust Head and Professor of the Internet of Things Thrust at the Hong Kong University of Science and Technology (Guangzhou), and an Emeritus Professor of the Department of Electronic and Computer Engineering at the Hong Kong University of Science and Technology.
\end{IEEEbiography}
\end{document}